\documentclass[paper]{JHEP3} 

\usepackage{epsfig,multicol,multirow}
\usepackage{amsmath,subfigure,latexsym,amssymb}
\usepackage{here}

\newcommand\fverb{\setbox\fverbbox=\hbox\bgroup\verb}
\newcommand\fverbdo{\egroup\medskip\noindent%
			\fbox{\unhbox\fverbbox}\ }
\newcommand\fverbit{\egroup\item[\fbox{\unhbox\fverbbox}]}
\newbox\fverbbox

\def\lsi{\raise0.3ex\hbox{$<$\kern-0.75em\raise-1.1ex\hbox{$\sim$}}}
\def\gsi{\raise0.3ex\hbox{$>$\kern-0.75em\raise-1.1ex\hbox{$\sim$}}}

\newcommand{\beq}{\begin{equation}}
\newcommand{\eeq}{\end{equation}}
\newcommand{\beqa}{\begin{eqnarray}}
\newcommand{\eeqa}{\end{eqnarray}}

\newcommand{\del}{\partial}
\newcommand {\tr}{{\rm tr\,}}

\newcommand{\al}{\alpha}
\newcommand{\be}{\beta}
\newcommand{\gm}{\gamma}

\newcommand{\ps}{\psi}

\newcommand{\oot}{\frac{1}{2}}

\newcommand{\oof}{\frac{1}{4}}

\newcommand{\intdt}{\int_0^\be \!dt\,}

\newcommand {\ee}{\mbox{e}}

\allowdisplaybreaks

\title{Nonperturbative studies of supersymmetric matrix quantum mechanics
with 4 and 8 supercharges\\
at finite temperature}

\author{Masanori Hanada,$^{a,b}$ So Matsuura,$^{c}$
Jun Nishimura$^{d,e}$ and Daniel Robles-Llana$^{a}$
\vspace*{0.5cm} \\
\llap{$^a$}Department of Particle Physics and Astrophysics,\\
Weizmann Institute of Science, Rehovot 76100, Israel\\
\llap{$^b$}Department of Physics, University of Washington, 
Seattle, WA 98195-1560, USA\\
\llap{$^c$}Department of Physics,
and Research and Education Center for Natural Science,\\
Keio University, 4-1-1 Hiyoshi, Yokohama, 223-8521, Japan\\
\llap{$^d$}KEK Theory Center,
High Energy Accelerator Research Organization (KEK),\\
Tsukuba, Ibaraki 305-0801, Japan\\
\llap{$^e$}Department of Particle and Nuclear Physics, 
School of High Energy Accelerator Science,\\
Graduate University for Advanced Studies (SOKENDAI),\\
Tsukuba, Ibaraki 305-0801, Japan
\vspace*{0.5cm} \\
\email{mhanada@u.washington.edu, s.matsu@phys-h.keio.ac.jp,\\
jnishi@post.kek.jp, daniel.robles@weizmann.ac.il}}

\preprint{WIS/17/10-DEC-DPPA\\ KEK-TH-1429}

\abstract{
We 
investigate thermodynamic properties of
one-dimensional U($N$) supersymmetric gauge
theories
with 4 and 8 supercharges
in the planar large-$N$ limit by Monte Carlo calculations. 
Unlike the 16 supercharge case,
the 
threshold bound state with zero energy
is widely believed not to exist in these models.
This led A.V.Smilga to conjecture that 
the internal energy decreases exponentially at low temperature
instead of decreasing with
a power law.
In the 16 supercharge case,
the latter behavior was predicted 
from the dual black 0-brane geometry
and confirmed recently by Monte Carlo calculations.
Our 
results for the models with 4 and 8 supercharges
indeed support the exponential behavior,
revealing a qualitative difference from the 16 supercharge case.
}


\begin{document}

\section{Introduction}

Large-$N$ supersymmetric gauge theories in low dimensions 
are useful laboratories to test the AdS/CFT correspondence 
and its generalizations \cite{Maldacena97,IMSY98}.
Among these, the 1d case is particularly simple since
the gauge theory is just a quantum mechanical system, 
and therefore one may hope to test the duality relations 
explicitly and to understand them more in depth.
Indeed Monte Carlo studies of 
such a system with
maximal supersymmetry (16 supercharges)
have been performed by using non-lattice 
simulations \cite{HNT07},
which reproduced various predictions from the dual string theory
including $\alpha'$-corrections
in the planar large-$N$ limit
\cite{AHNT07,HMNT08,HHNT08,Hanada:2009ne}.\footnote{A lattice 
simulation of the
same model has been performed in refs.~\cite{CW07,CW08}
with qualitatively consistent results 
for thermodynamical quantities.
See also refs.~\cite{KLL} for earlier calculations based
on the Gaussian 
expansion method.
}
Another reason for interest in the one-dimensional case
is that
the same model, which represents the worldvolume theory 
of $N$ D0-branes, has been proposed as a fully non-perturbative
formulation of uncompactified M-theory 
in the light-cone frame \cite{BFSS96}.
Interesting observations relevant in this direction
are also obtained from
Monte Carlo calculations of correlation functions \cite{Hanada:2009ne}.

In this paper we apply the same Monte Carlo method
to the study of non-maximally supersymmetric quantum
mechanics with 4 and 8 supercharges.\footnote{In fact
the 4 supercharge model has been studied \cite{HNT07,CW07}
prior to the 16 supercharge model for the purpose of testing 
the method. 
In order to address the issues given below, however,
we need to study the system at much lower temperature,
as we do in this paper.
}
As opposed to the model with 16 supercharges, there is no
known gravity dual for these models.
On the other hand,
a property common 
to
all these supersymmetric models
is that there exist flat directions in the potential,
which are not lifted quantum mechanically 
due to supersymmetry
unlike their
bosonic counterparts. 
As a result, the theory contains
not only the discrete states that the bosonic models have,
but also the scattering states forming
the continuous branch of the spectrum \cite{Smilga:1984jg}.
In the 16 supercharge model,
on top of the states just mentioned,\footnote{A detailed analysis
of the continuum spectrum and its implications on the physics
of supermembranes have been given in refs.~\cite{de Wit:1988ct}
and \cite{Smilga:1989ew}.
}
it is known that there exists
a threshold bound state, which is 
somewhat extended in the flat directions
and yet has a finite norm \cite{Yi:1997eg,Sethi:1997pa,Moore:1998et}.
Such a state, 
which is crucial for the M-theory interpretation \cite{BFSS96}, 
is considered not to exist
in the non-maximally supersymmetric 
models \cite{Yi:1997eg,Sethi:1997pa,Moore:1998et}.

Recently A.V.Smilga \cite{Smilga:2008bt} conjectured that
the above difference between the 16 supercharge model
and the other supersymmetric models 
may lead to a qualitative difference
in the temperature dependence of the internal energy.
In the 16 supercharge case the power-law behavior $E \propto T^{14/5}$
at low $T$ 
was predicted from the dual black 0-brane geometry \cite{IMSY98},
and was confirmed by the 
Monte Carlo simulation \cite{AHNT07,HHNT08,CW08}.
Smilga first gave a theoretical argument on the gauge theory side
for the particular power ``14/5''.
Here an important role is played by
normalizable excitations around
the threshold bound state
with a new energy scale (proportional to $N^{-5/9}$)
suggested from the effective Hamiltonian 
for the relevant O($N$) degrees of freedom 
in the flat directions.
For the 4 and 8 supercharge cases, he conjectured that
the internal energy decreases exponentially 
$E \propto \exp(-c/T)$ at low $T$
assuming the absence of normalizable states
with a new energy scale.
However, he also mentioned a possibility that
there exist normalizable states 
associated with the effective Hamiltonian
with a new energy scale proportional to $N^{-1}$. 
In that case one obtains $E \propto T^2$ at low $T$
following the same argument as in the 16 supercharge case.
Our Monte Carlo data 
support the exponential behavior
rather than the power-law behavior.
%
%

As other basic properties of the supersymmetric matrix quantum mechanics,
we also study
the phase structure along the temperature axis,
which turns out to be qualitatively the same for all the supersymmetric models.
There is only one phase, in which 
the center symmetry is broken spontaneously.


The rest of this paper is organized as follows.
In section \ref{sec:model} we define the supersymmetric matrix quantum
mechanics
and discuss their basic properties.
In section \ref{sec:thermodynamics} we briefly review Smilga's argument.
In section \ref{sec: finite temp numerics} we present our Monte Carlo 
results for the internal energy and compare them with
the low temperature behaviors suggested by Smilga.
In section \ref{sec:su-two} we compare our results
with the energy spectrum obtained in the 4 supercharge model
for the SU(2) gauge group \cite{Campostrini:2004bs}.
Section \ref{sec:summary} is devoted to a summary and discussions. 
In appendix \ref{sec:EO} we derive the expression we use to
calculate the internal energy in actual Monte Carlo simulation.

\section{The models and their basic properties}
\label{sec:model}

The supersymmetric matrix quantum mechanics
are
defined by the action
\beqa
\label{action}
S = \frac{1}{g^2} \intdt {\rm tr}
 \left\{
 \oot (D_t X_i)^2 
+\oot \psi_\alpha D_t \psi_\alpha
-\oof [X_i,X_j] ^2
- \oot \ps_\alpha (\gm_i)_{\alpha\beta} [X_i , \ps_\beta ]
\right\} \ ,
\eeqa  
where
$D_t \equiv \partial_t-i[A(t),\,\,\cdot\,\,]$ 
represents the covariant derivative.
The bosonic matrices
$A(t)$, $X_i(t)$ $(i=1,2,\cdots,d)$ 
and the fermionic matrices
$\ps_\al(t)$ $(\al=1,2,\cdots,p)$ are 
$N \times N$ Hermitian matrices,
where $p=4,8,16$ for $d=3,5,9$, respectively.
The models can be obtained formally
by dimensionally reducing
${\cal N}=1$ super Yang-Mills theory
in $D=d+1$ dimensions to one dimension,
and they can be viewed as a 1d
U($N$) gauge theory, where
$A(t)$, $X_i(t)$ and $\ps_\al(t)$
are the gauge field, adjoint scalars and 
spinors, respectively.\footnote{Here 
we use the notation of Majorana
spinors to describe the $D=4,6,10$ cases in a unified manner.
When we write a code for the $D=4,6$ models, we 
rewrite the action 
in terms of Weyl fermions, which has only $\frac{p}{2}$ components.
} 
The $p \times p$ symmetric matrices 
$\gm_i$ obey the Euclidean Clifford algebra
$\{ \gm_i,\gm_j \}=2\delta_{ij}$.
The models are supersymmetric,
and the number of supercharges is given by $p$.

In order to study
the thermodynamics,
we impose 
periodic and anti-periodic boundary conditions
on the bosonic and fermionic matrices,
respectively, which breaks supersymmetry.
The extent $\beta$ in the Euclidean time direction
$t$ then corresponds to the inverse
temperature $\beta=T^{-1}$.

The action is invariant under the shifts
\beq
A(t)  \mapsto  A(t) + \alpha(t) {\bf 1} \ , 
\quad
X_i(t)  \mapsto  X_i (t) + x_i {\bf 1} \ ,
\eeq
where $\alpha(t)$ is an arbitrary periodic function,
and $x_i$ 
is an arbitrary constant.
In order to remove the corresponding decoupled modes,
we impose the conditions
\beqa 
\label{czm}
\tr  A(t) =0 \ , \quad
\intdt \tr  X_i (t) = 0   
\eeqa
for all $i=1 , 2 , \cdots , d$.

%
Since the coupling constant $g$ can be 
absorbed by rescaling the matrices and $t$
appropriately, we set the 't~Hooft coupling constant
$\lambda \equiv g^2 N$ to unity
without loss of generality.
This implies that we replace the prefactor $\frac{1}{g^2}$
in the action (\ref{action}) by $N$ in what follows.
In order to 
put the theory on a computer \cite{HNT07},
we first fix the gauge as
$A(t)=\frac{1}{\beta}{\rm diag}(\alpha_1,\cdots,\alpha_N)$,
where $-\pi<\alpha_i\le \pi$, 
include the corresponding Fadeev-Popov determinant,
and then
introduce a Fourier mode cutoff $\Lambda$ as
\beq
X_i(t) = \sum_{n=-\Lambda}^\Lambda\tilde{X}_{in}e^{i n \omega t}  \ ,
\quad 
\psi_\alpha(t) =
\sum_{r=-\Lambda'}^{\Lambda'}
\tilde{\psi}_{\alpha n}e^{i r \omega t} \ ,
\label{Fourier-expand}
\eeq
where $\omega=\frac{2\pi}{\beta}$.
The indices $n$ and $r$ run over integers
and half integers, respectively,
and we set $\Lambda' = \Lambda - \frac{1}{2}$.
%
The breaking of supersymmetry due to finite $\Lambda$
is shown to disappear quickly with increasing $\Lambda$ \cite{HNT07}.
The fermion determinant is positive semi-definite for 
the $D=4$ model 
\footnote{The proof goes similarly
to the case of the totally reduced 
model \cite{AABHN00}.} 
even at finite $\Lambda$.
However, it is generally complex for the $D=6$ and $D=10$ models.
As is done in previous works for the $D=10$ model,\footnote{Previous 
results for the $D=10$ model obtained in this way
agreed well with the predictions from 
the gauge-gravity duality \cite{AHNT07,HMNT08,HHNT08,Hanada:2009ne}.
It is therefore expected that the fluctuation of the phase is 
totally decorrelated with all physical quantities.
While there is some evidence that this is indeed the case,
complete understanding is still missing.}
we simply omit the phase of the fermion determinant
when we study the $D=6$ model.
Simulations has been performed 
with the Rational Hybrid Monte Carlo (RHMC) algorithm \cite{RHMC},
which is quite standard in recent lattice QCD simulations.

As a fundamental quantity in thermodynamics,
the free energy 
${\cal F} = - \frac{1}{\beta} \ln Z(\beta)$
is defined in terms of the partition function
\beq
Z(\beta) = \int 
[{\cal D} A]_\beta
[{\cal D} X]_\beta  [{\cal D} \psi]_\beta 
 \, \ee^{- S(\beta)} \ ,
\label{def-part-fn}
\eeq
where the suffix of the measure $[ \ \cdot  \ ]_\beta $
represents the period of the field
to be path-integrated.
However, the evaluation of the partition function 
$Z(\beta)$ is not straightforward in 
Monte Carlo simulation.
We therefore study the internal energy defined by
\beq
E \equiv \frac{d}{d \beta} (\beta {\cal F}) 
  = - \frac{d}{d \beta} \log Z(\beta) \ ,
\label{defE}
\eeq
which has equivalent information as 
the free energy,
given the boundary condition ${\cal F} = E$ at $T=0$.
In Appendix \ref{sec:EO}
we explain how we calculate the internal
energy $E$ in actual simulation.




A common property 
of the supersymmetric matrix quantum mechanics
for all $D=4,6,10$,
which distinguishes them from the bosonic counterparts,
is that they have a continuum branch of the energy 
spectrum associated with the flat directions $[X_i , X_j] = 0$
in the potential.
Its existence can be seen in Monte Carlo simulations 
as a run-away behavior, which can be probed by
the observable
\beq
\label{def R2}
R^2 \equiv
\frac{1}{N\beta} \intdt \tr \Bigl( X_i(t) \Bigr)^2 \ .
\eeq
In the high-$T$ limit the fermions decouple, and the system
becomes essentially a bosonic model, which is well-defined for any $N \ge 2$.
The expectation value $\langle R^2 \rangle $ at large but finite $T$
can be obtained without any problem,
and it can be nicely reproduced
by the high temperature expansion 
including the subleading term \cite{Kawahara:2007ib}.
As we lower $T$ for a fixed $N$, 
the expectation value $\langle R^2 \rangle$ 
decreases in accord with the high temperature expansion.
%
However, at some $T$, we observe
some cases in which the value of $R^2$ 
starts to increase endlessly 
during the simulation, which represents the aforementioned run-away behavior.
%
If we use larger values of $N$, we can go to lower $T$ without
seeing such a behavior.
Theoretical understanding of this property
is provided in ref.~\cite{AHNT07} based on the one-loop effective action.
We therefore consider that a well-defined ensemble 
can be defined at any finite $T$
in the large-$N$ and large-$\Lambda$ limits,
and that the ensemble average corresponds to taking the expectation value
within the Hilbert space restricted to the normalizable states.

In fig.~\ref{fig:energy_trx}
we plot the expectation value $\langle R^2 \rangle$ 
obtained by Monte Carlo simulation. 
At low $T$ and for $N$ not extremely large,
we find some cases in which $\langle R^2 \rangle$ increases rapidly
as the cutoff $\Lambda$ is increased.
This can be understood as finite $\Lambda$ effects,
which tend to lift the flat directions slightly \cite{AHNT07},
and hence to suppress the run-away behavior.
One can also see from fig.~\ref{fig:energy_trx}
that larger $N$ tends to suppress the run-away behavior.

\FIGURE[t]{
    \epsfig{file=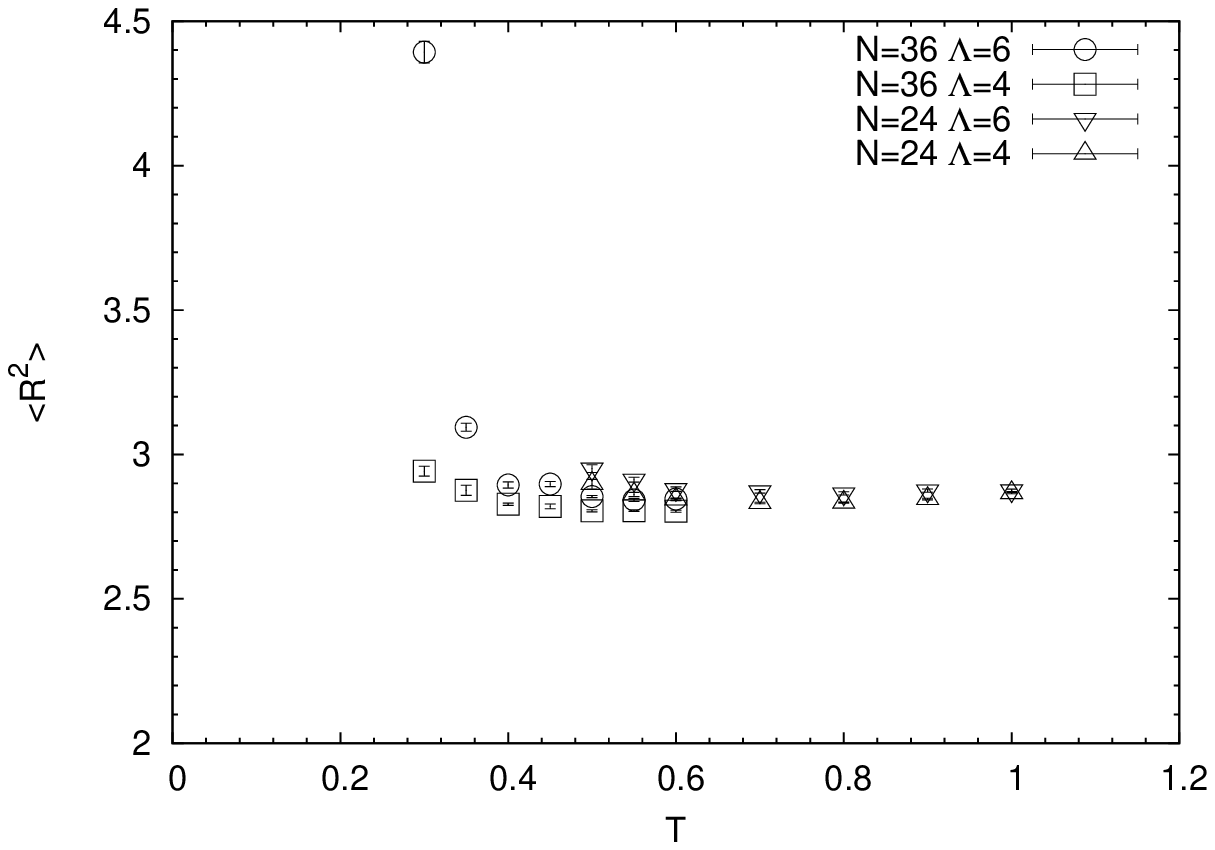,%
width=7.0cm}
    \epsfig{file=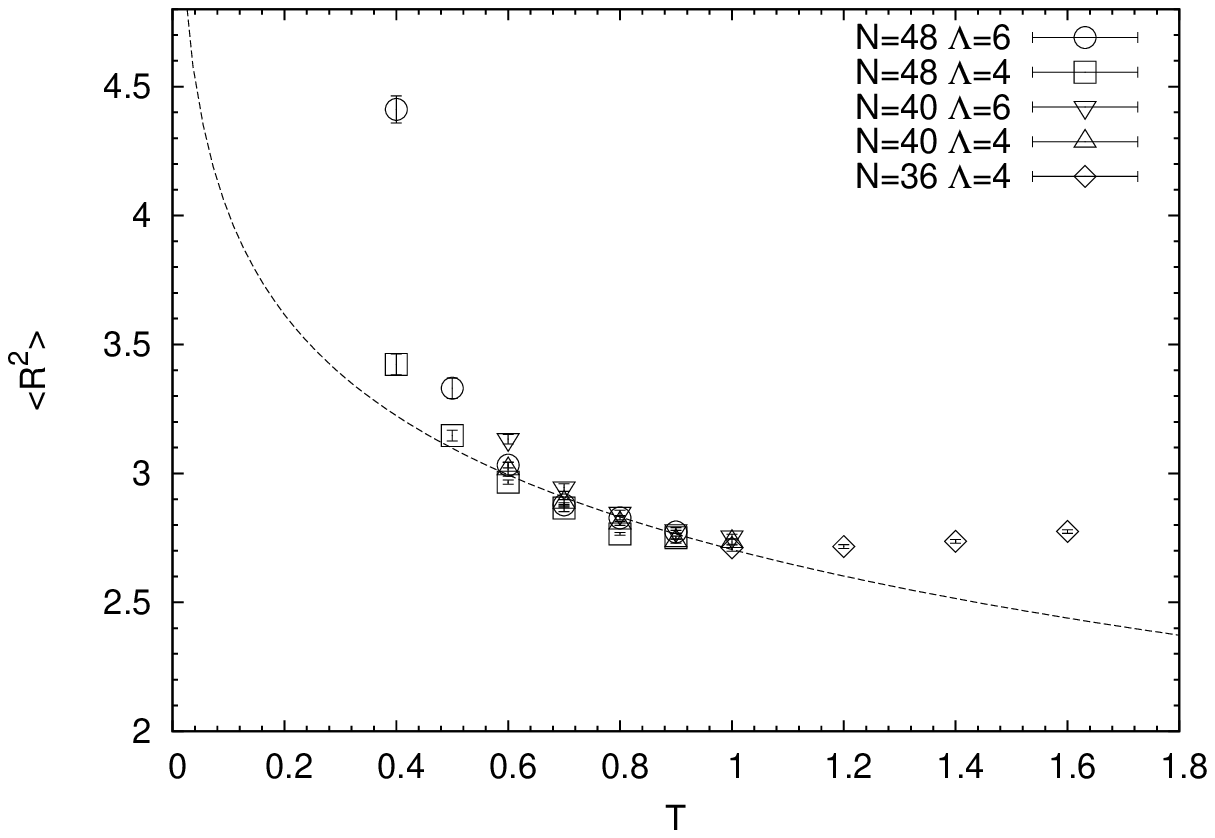,%
width=7.0cm}
\caption{The extent of space is plotted against the temperature
for the $D=6$ model (left) and for the
$D=4$ model (right).
The dashed line in the right panel represents 
a possible log-divergent behavior $\langle R^2 \rangle 
= a + b \log (1/T)$, where $a=2.71$ and $b=0.566$
obtained by fitting the data for
$N=48$, $\Lambda=6$, $T=0.6, \cdots , 0.9$.
}
\label{fig:energy_trx}
}

Our data suggest that
the value of $\langle R^2 \rangle$ in the above limit
behaves differently for the $D=6$ and $D=4$ models.
For $D=6$, it decreases monotonically (or stays almost constant) 
as $T$ decreases 
similarly to what is observed for $D=10$ \cite{AHNT07}.
For $D=4$, on the other hand, 
it starts to increase as $T$ is lowered below $T \sim 1$.
This behavior is reminiscent of
the divergence of the second moment in the totally reduced model
of 4d ${\cal N}=1$ super Yang-Mills theory \cite{KS99,AABHN00}.
Since fermions obey anti-periodic boundary conditions in our 1d model, 
the temperature $T$ plays a role of the SUSY breaking mass parameter
for the fermions.
This gives a regularization to the second moment $\langle R^2 \rangle$,
which is logarithmically divergent with respect to the regularization
parameter.
Hence we obtain $\langle R^2 \rangle \sim \log \frac{1}{T}$.
Our results in fig.~\ref{fig:energy_trx} (right)
are consistent with this behavior, but more data with larger $N$
around $T\sim 0.4$ are needed to confirm it unambiguously.


As another property of the supersymmetric matrix quantum mechanics,
let us discuss the phase structure along the temperature axis.
For that we define the Polyakov line
\beq
\label{def P}
P \equiv
\frac{1}{N}\tr 
\mathcal{P} \exp
\left(
i\intdt A(t)
\right) \ ,
\eeq
where the symbol ``$\mathcal{P} \exp$'' 
represents the path-ordered exponential.
It serves as the order parameter of the spontaneous breaking of
the center symmetry.
In fig.~\ref{fig:Pol} we plot the expectation value
$\langle |P| \rangle$.
In both $D=6$ and $D=4$ models,
the results can be fitted by 
$\langle |P| \rangle \sim a\exp\Bigl(-\frac{b}{T}\Bigr)$,
which is a characteristic low-$T$ behavior 
of a ``non-confining'' theory.
This implies that the center symmetry is broken
at any finite temperature
similarly to the case of maximal supersymmetry \cite{AHNT07}.

The bosonic models are known to undergo a ``deconfining transition''
at a critical temperature \cite{latticeBFSS,KNT07,Mandal:2009vz}, 
below which the center symmetry remains unbroken.\footnote{In fact
there are two phase transitions along the temperature axis 
as shown by Monte Carlo simulation \cite{KNT07}
and by $1/D$ expansion \cite{Mandal:2009vz}.}
Therefore, the Eguchi-Kawai equivalence \cite{EK82}
(or the ``volume independence'' \cite{NN03}) holds
above the critical $\beta$ in the bosonic models.
As a result, the internal energy does not depend on $T$
below some critical value.
The physical meaning of this behavior is discussed in the next section.

\FIGURE{
    \epsfig{file=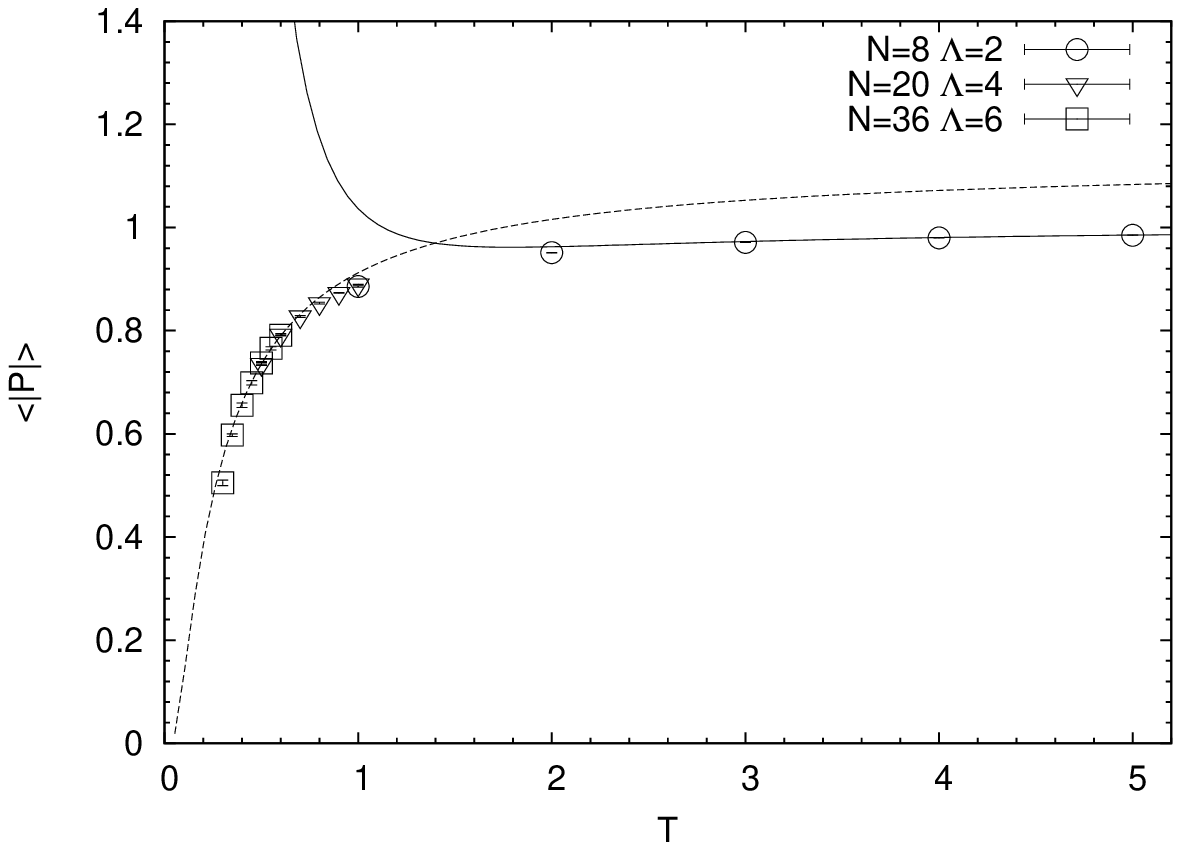,%
width=7.0cm}
    \epsfig{file=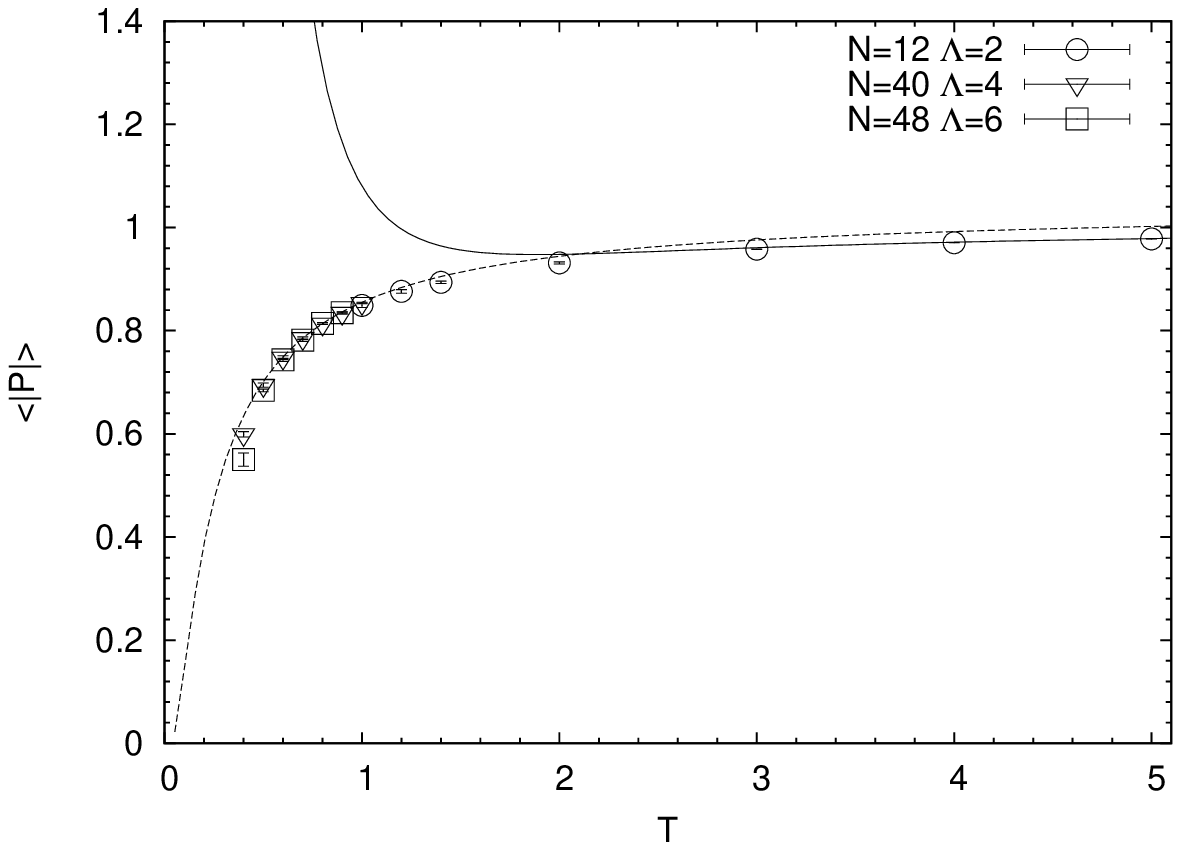,%
width=7.0cm}
\caption{The expectation value $\langle |P| \rangle$
is plotted against the temperature
for the $D=6$ model (left) and for the
$D=4$ model (right).
The solid line represents the results 
of the high temperature expansion \cite{Kawahara:2007ib}
(up to the next leading order) 
at $N=8$ for $D=6$ and at $N=12$ for $D=4$,
and the dashed line represents a fit to the behavior 
$\langle |P| \rangle = a\exp(-\frac{b}{T})$
characteristic to a ``non-confining'' theory.
The parameters obtained by fitting are
$a=1.13(1)$ and $b=0.215(6)$ for the $D=6$ model
and $a=1.03(1)$ and $b=0.19(1)$ for the $D=4$ model.
}
\label{fig:Pol}
}

\section{Brief review of Smilga's argument}
\label{sec:thermodynamics}

In this section we review Smilga's argument \cite{Smilga:2008bt}
for the low-$T$ behavior of the internal energy.
Note that the characteristic energy scale of 
the matrix quantum mechanics
is $E_{\rm char} \sim \lambda^{\frac{1}{3}}$, 
which is O(1) in our convention $\lambda=1$ .



Let us first start with the bosonic models, which can be obtained
by omitting the fermions.
The large-$N$ behavior of such models 
was studied by Monte Carlo simulation in ref.~\cite{KNT07}.
The normalized internal energy $\frac{1}{N^2} E$
as a function of $T$
is found to stay 
constant
$\frac{1}{N^2} E = \varepsilon_0$
below the
critical temperature.
This can be viewed as a consequence of the Eguchi-Kawai equivalence
mentioned in the previous section.
Here $E_{\rm vac} \equiv N^2 \varepsilon_0$ can be identified
as the vacuum energy of the bosonic model.\footnote{For instance, 
$\varepsilon_0 = 6.695(5)$ is obtained for the
$D=10$ bosonic model by Monte Carlo simulation \cite{KNT07}.
Ref.~\cite{Smilga:2008bt} gives an explanation for
$E_{\rm vac} \sim {\rm O}(N^2)$ based on the variational method.
} 
Beyond the critical temperature,
the normalized internal energy $\frac{1}{N^2} E$
starts to grow with $T$.
This behavior of $E(T)$ can be understood from the behavior of the
entropy $S(E)$ using the relations
\beq
Z = \int dE \, \ee^{S(E)} \ee^{-\beta E} 
\label{partition-S-E}
\eeq 
and $E = - \frac{\del}{\del \beta}\log Z$.
The growth of the entropy $S(E)$ changes
qualitatively\footnote{The 
phase transition is found to be 
of second order for $D=10$ \cite{Mandal:2009vz}.
This implies that $\frac{dS(E)}{dE}$ is continuous
but $\frac{d^2 S(E)}{dE^2}$ is discontinuous at $E=E_{\rm cr}$.
} 
at some critical energy
$E_{\rm cr}$, where $E_{\rm cr} - E_{\rm vac} = {\rm O}(1)$.
Since the region  $E < E_{\rm cr}$ corresponds to the confined phase,
the entropy is $S(E)\sim {\rm O}(1)$.
It is then clear that $\frac{1}{N^2} E = \varepsilon_0$ 
below some critical temperature.
As the energy approaches the critical value $E_{\rm cr}$,
the entropy $S(E)$ grows rapidly and becomes O$(N^2)$.
(This transition has been studied
in refs.~\cite{Furuuchi:2003sy,Aharony}.)
Since the region  $E > E_{\rm cr}$ corresponds to the deconfined phase,
it is more appropriate to describe
the system at large $N$
in terms of the normalized entropy $\sigma \equiv \frac{1}{N^2}S$
as a function of the normalized energy $\varepsilon \equiv \frac{1}{N^2}E$.
In particular, the saddle-point approximation becomes exact
in evaluating (\ref{partition-S-E}) at $N=\infty$, and one obtains 
\beq
\frac{{\rm d} \sigma}{{\rm d} \varepsilon} = \beta \ ,
\label{sigma-beta}
\eeq
which gives the normalized internal energy $\varepsilon$
as a function of the temperature.
This explains why $\varepsilon$ has nontrivial
dependence on $\beta$
above some critical temperature.


Next we discuss the case of supersymmetric matrix quantum mechanics.
Here we consider only the normalizable states, which are
included in our Monte Carlo simulation as we discussed
below eq.~(\ref{def R2}).
Since the supersymmetric models are not confining,
it is more appropriate to describe
the system at large $N$
in terms of the normalized entropy $\sigma \equiv \frac{1}{N^2}S$
as a function of the normalized energy $\varepsilon \equiv \frac{1}{N^2}E$.

For the $D=10$ model one obtains
\beq
\varepsilon \sim  7.41 \, T^{\frac{14}{5}} \ 
\label{power-lawD10}
\eeq 
at low $T$
from the gauge-gravity duality, and this power-law behavior 
(including the coefficient) is
confirmed accurately by Monte Carlo simulation\footnote{The power 
of the
subleading term at low $T$ was also derived in ref.~\cite{HHNT08} 
from the gravity side by considering the $\alpha '$ corrections. 
Monte Carlo results 
reproduced that power, too.
} \cite{HHNT08}.
From eq.~(\ref{sigma-beta}), one readily finds that
\beq
\sigma \sim \varepsilon ^{p} 
\label{sigma-eps}
\eeq
for small $\varepsilon$ with the power 
\beq
p=\frac{9}{14} \ .
\label{power-p}
\eeq


Rewriting (\ref{sigma-eps})
in terms of unnormalized variables, we obtain
\beq
 S \sim N^2 \left(\frac{E}{N^2} \right) ^{p}  
\label{power-S-E}
\eeq
for small $\frac{1}{N^2}E$.
If we assume, for simplicity, that this formula holds even 
at $E \sim {\rm O}(1)$,
we obtain $S \sim {\rm O}(N^{2(1-p)})$,
which suggests 
that there are many normalizable states with 
the energy $E \sim {\rm O}(1)$.
(We will shortly make this argument more precise.)
In ref.~\cite{Smilga:2008bt} it was speculated that
these states should be 
related to 
the existence of the threshold bound state in the $D=10$ model.

Indeed
one can reproduce the power (\ref{power-p})
by considering the low energy excitations around the threshold bound state.
Let us first recall that
the supersymmetric models can
have states extended in the flat directions.
Wavefunctions for such states can be written symbolically
as
\begin{equation} 
\Psi
=\chi (x_{\rm slow}) \, \psi(x_{\rm fast}) \ ,
\label{BOapprox}
\end{equation}
using the Born-Oppenheimer approximation.
Here $x_{\rm slow}$ parametrizes the slow oscillations along the flat
directions corresponding to the Cartan subalgebra of the gauge group, 
whereas $x_{\rm fast}$ parametrizes 
the fast oscillations along the orthogonal directions.
Typically these states correspond to non-normalizable states,
and the spectrum becomes continuous.
In the $D=10$ case, however, there also
exists a threshold bound state,
which is a normalizable state with 
zero energy that can be expressed in the form (\ref{BOapprox}).
The excitations around such a state is expected to 
have a new energy scale $E_{\rm char}^{\rm new} \sim N ^{-\frac{5}{9}}$ 
\cite{Smilga:2008bt}
from the form of 
the effective Hamiltonian \cite{Becker:1997wh,Okawa:1998pz}
for the states (\ref{BOapprox}).
Considering that the degrees of freedom 
in the effective Hamiltonian is ${\rm O}(N)$,
the entropy should be $S(E) = {\rm O}(N)$ for
the energy $E \sim N E_{\rm char}^{\rm new} = {\rm O}(N^{\frac{4}{9}})$.
Imposing this requirement\footnote{\label{diff-Smilga}In 
ref.~\cite{Smilga:2008bt} the power
was determined by matching the energy at temperature
$T \sim E_{\rm char}^{\rm new}$ with
$E \sim N E_{\rm char}^{\rm new}$, which is equivalent to the argument 
given here.
}
on the power law behavior (\ref{power-S-E}),
one can determine the power $p$ and finds that it 
agrees precisely with (\ref{power-p}).

Strictly speaking, 
the prediction (\ref{power-lawD10})
from the gauge-gravity duality is valid 
for $ N^{-\frac{10}{21}} \ll T \ll 1$.
On the other hand, here we are considering an energy scale
$E \sim  {\rm O}(N^{\frac{4}{9}})$, which corresponds to the
temperature $T\sim {\rm O}(N^{-\frac{5}{9}})$
slightly below the lower bound of the validity region. 
The precise agreement therefore implies that the power law
(\ref{power-lawD10})
actually holds for a range of $T$ wider than naively expected.
Similar observation was made in the comparison of
correlation functions with predictions from the gauge-gravity 
duality \cite{Hanada:2009ne}.
The power-law behavior of the correlation functions
obtained from the gauge-gravity duality
was found to hold in the far infrared regime, in which the
supergravity analysis is not valid naively.

For the $D=4,6$ cases, calculation of the Witten index suggests
that the threshold bound state does not 
exist \cite{Yi:1997eg,Sethi:1997pa,Moore:1998et}.
Therefore one may naively consider that there is no normalizable state
written in the form (\ref{BOapprox}).
The lowest energy level\footnote{In the $D=4$ model with $N=2$,
it was shown explicitly that these states form a 
supermultiplet \cite{Campostrini:2004bs} and has a non-zero energy.
See section \ref{sec:su-two} for more details.} in 
the discrete spectrum is considered to 
have the energy $E_{\rm vac}= {\rm O}(1)$,
and hence we have
$\varepsilon \rightarrow 0$ as $T \rightarrow 0$ in the large-$N$ limit.
Then the entropy is $S(E)\sim {\rm O}(1)$ for 
$E\sim {\rm O}(1)$, and it is expected to increase 
almost linearly with the energy $E$
as it approaches $S(E)\sim {\rm O}(N^2)$ at $E\sim {\rm O}(N^2)$.
Therefore a more plausible behavior for the normalized entropy 
$\sigma(\varepsilon)$ at small $\varepsilon$ is 
\beq
\sigma \sim  \varepsilon   \log \left(\frac{1}{\varepsilon}  
\right)  \ ,
\label{sigma-linear}
\eeq
where the logarithmic factor is introduced to 
ensure $\varepsilon \rightarrow 0$ as $T\rightarrow 0$.
Then from eq.~(\ref{sigma-beta}) we obtain
\beq
\varepsilon \sim \ee^{-\frac{c}{T}} \ .
\label{exp-ansatz}
\eeq
On the other hand, 
the form of the effective Hamiltonian for the
continuum states (\ref{BOapprox})
suggests a new scale
$E_{\rm char}^{\rm new} \sim N^{-1}$ \cite{Smilga:2002mra}.
If there exist normalizable states 
with this new scale,
the same argument\footnote{Here it is assumed that
the form (\ref{power-S-E}) holds even at the energy scale
$E \sim N E_{\rm char}^{\rm new} = {\rm O}(1)$.
} 
as the $D=10$ case suggests (\ref{power-S-E})
with $p=\frac{1}{2}$, and therefore
\beq
\varepsilon
\sim T^{2} \ .
\label{power-D46}
\eeq
We will see that our Monte Carlo data support
(\ref{exp-ansatz}) rather than (\ref{power-D46}).

\FIGURE[t]{
    \epsfig{file=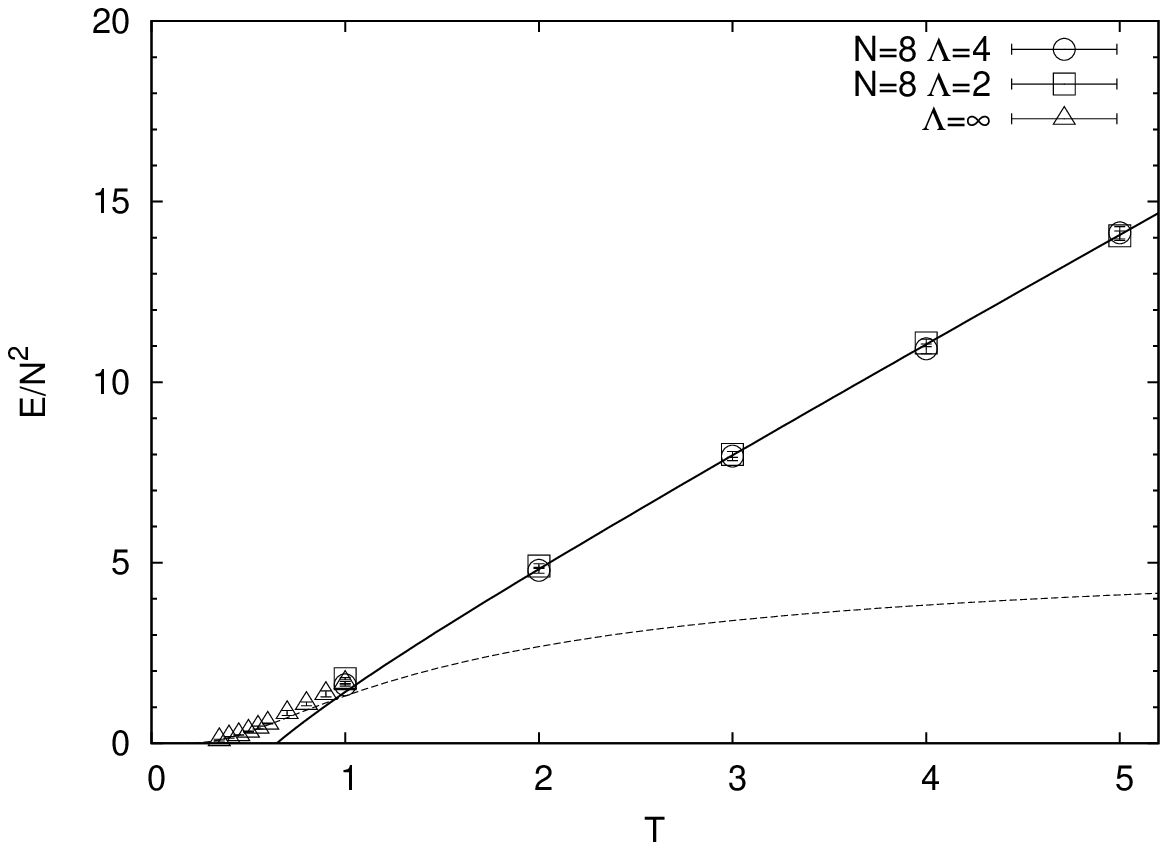,%
width=7.0cm}
    \epsfig{file=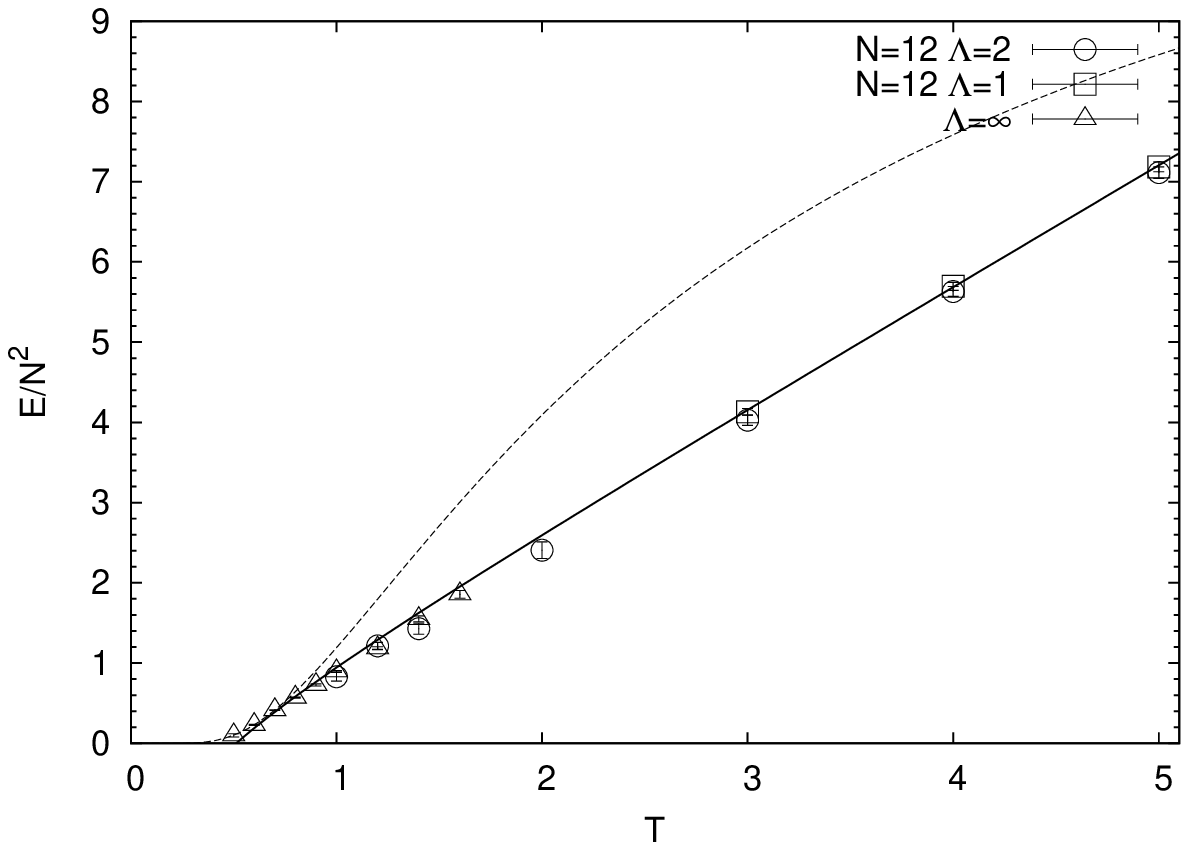,%
width=7.0cm}
\caption{
The normalized internal energy $\frac{1}{N^2}E$ is
plotted against temperature
for the $D=6$ model (left)
and the $D=4$ model (right).
The solid line represents the results 
of the high temperature expansion \cite{Kawahara:2007ib}
(up to the next leading order) 
at $N=8$ for $D=6$ and at $N=12$ for $D=4$.
The triangles represent the results obtained at low $T$ by
the large-$\Lambda$ extrapolation in fig.~\ref{fig:energy_cut}.
The dashed lines represent the behavior
(\ref{exp-ansatz-unnorm}) obtained by fitting the data at low $T$.
}
\label{fig:energy_HTE}
}

\FIGURE[t]{
    \epsfig{file=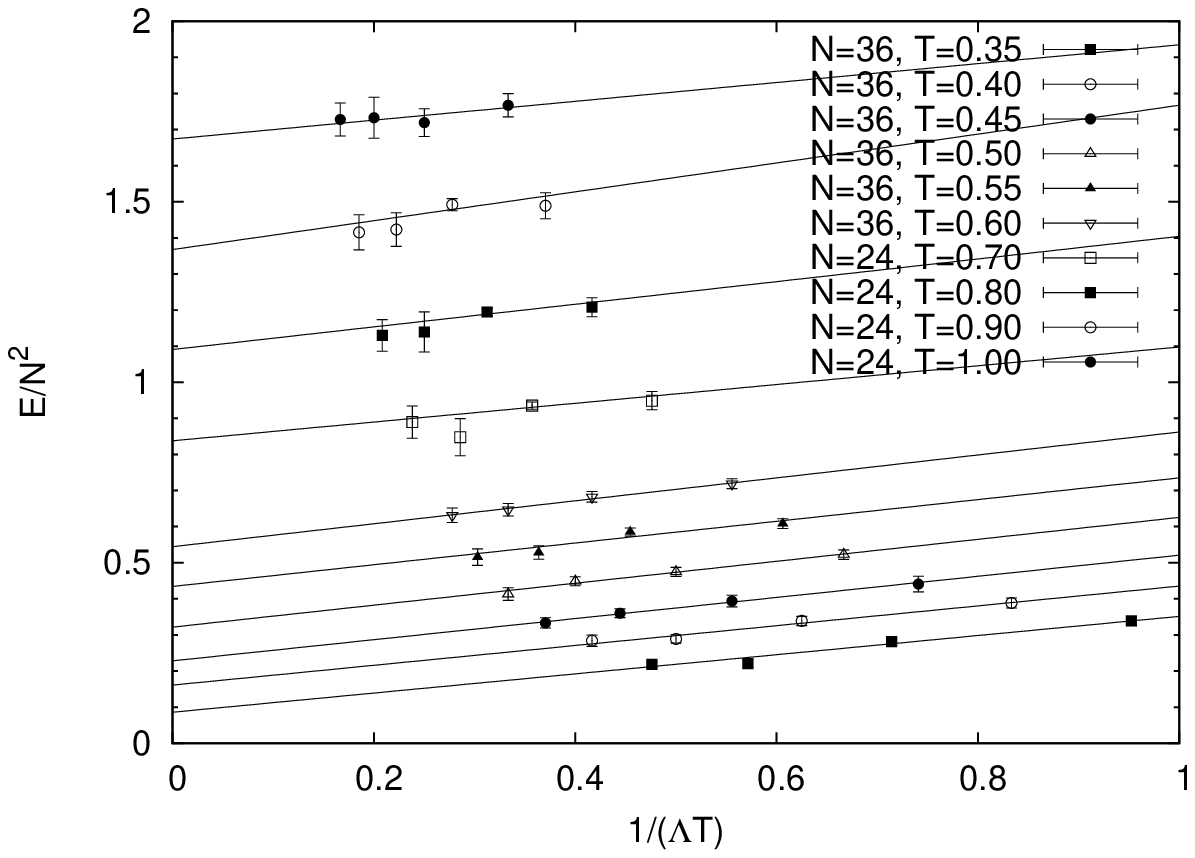,%
width=7.0cm}
    \epsfig{file=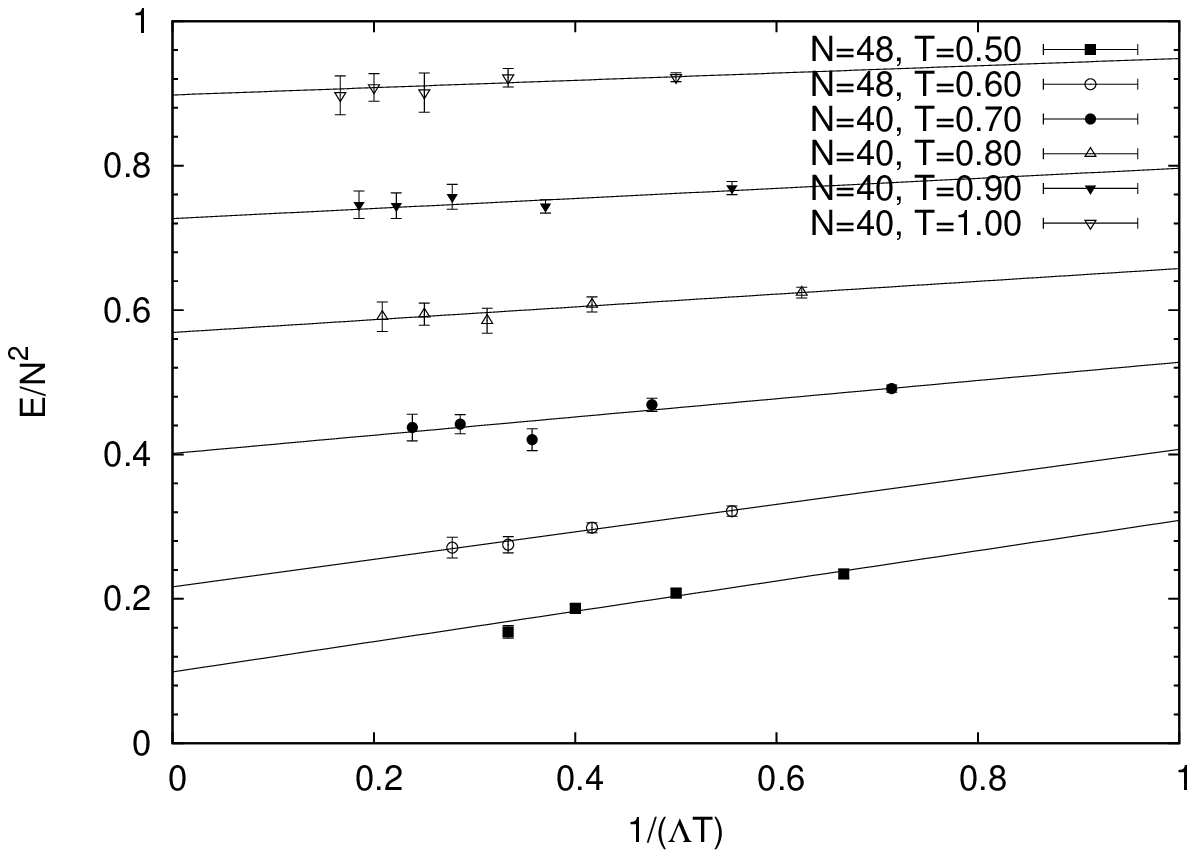,%
width=7.0cm}
\caption{The normalized energy $\frac{1}{N^2}E$ is plotted against 
$\frac{1}{\Lambda T}$
for the $D=6$ model (left) and for the $D=4$ model (right).
The data points for each $T$ can be fitted nicely to a straight line, 
which enables a sensible large $\Lambda$ extrapolation.
}
\label{fig:energy_cut}
}

\section{Results for the internal energy}
\label{sec: finite temp numerics}


In fig.~\ref{fig:energy_HTE}
we plot the internal energy at high temperature.
The results agree nicely with the results of 
the high temperature expansion obtained 
in ref.~\cite{Kawahara:2007ib}.

Next we consider the low-$T$ behavior.
Since the physical cutoff scale is given by
$\omega \Lambda =2 \pi T \Lambda $, which is
proportional to $\Lambda T$,
one needs to increase $\Lambda$
as one goes to lower $T$.
In fig.~\ref{fig:energy_cut} we plot the internal energy
against $\frac{1}{\Lambda T}$ for various $T$. 
We see that the data points for fixed $T$
can be fitted nicely by a straight line.
Based on this plot, we make an extrapolation to $\Lambda=\infty$.

Figure~\ref{fig:energy_log} 
shows the log plot
of the normalized internal energy extrapolated to $\Lambda =\infty$ 
as a function of the temperature for the $D=6,4$ models.
A straight line in this figure represents the behavior
\beq
\frac{1}{N^2} E = a \exp \left( -\frac{b}{T} \right) \ .
\label{exp-ansatz-unnorm}
\eeq
\FIGURE[t]{
    \epsfig{file=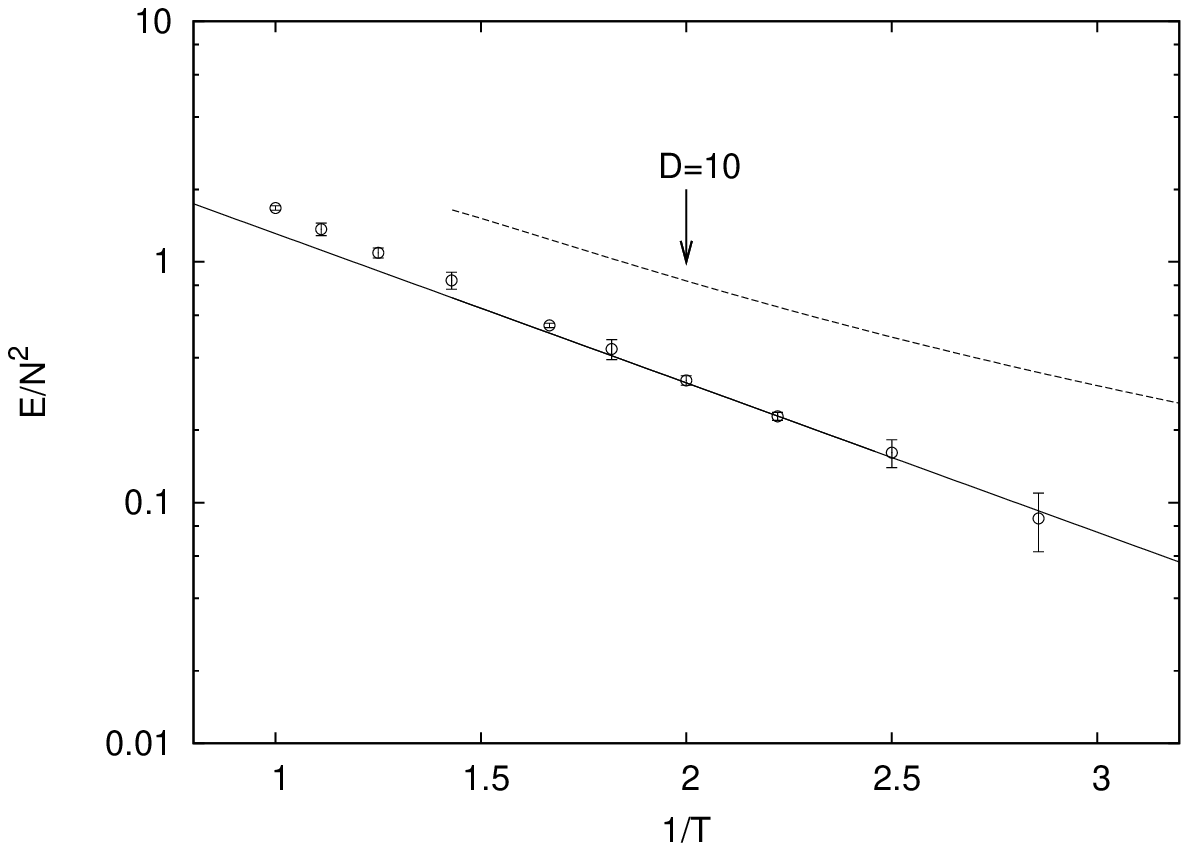,%
width=7.0cm}
    \epsfig{file=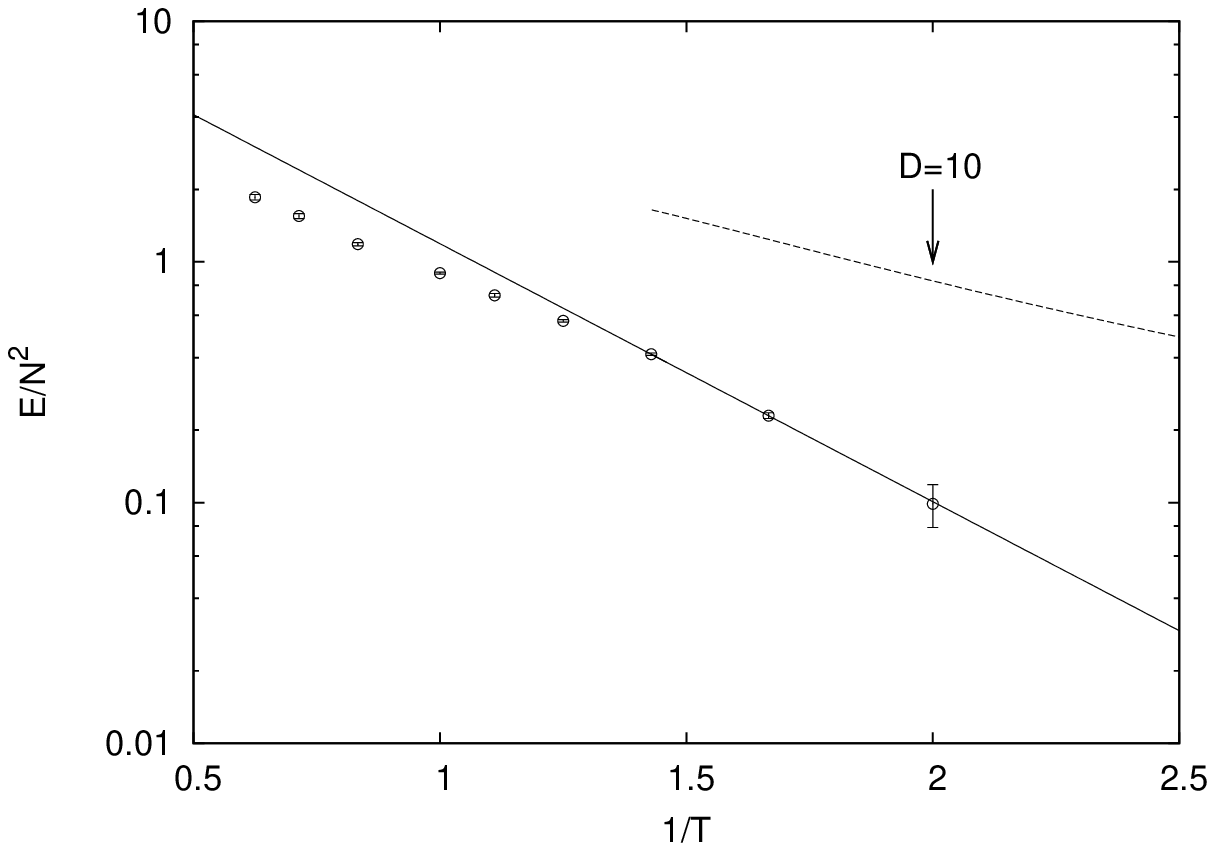,%
width=7.0cm}
\caption{The log plot of the
internal energy $\frac{1}{N^2}E$ against the temperature
for the $D=6$ model (left) and for the $D=4$ model (right).
The solid line represents a fit to the exponential behavior
(\ref{exp-ansatz-unnorm}) using the 3 data points with lowest $T$.
The dashed line represents the results for the $D=10$ model \cite{HHNT08}
obtained by fitting the Monte Carlo data to the form
$\frac{1}{N^2}E = 7.41 T^{\frac{14}{5}} 
- c T^{\frac{23}{5}}$, where $c=5.58$.
}
\label{fig:energy_log}
}
\FIGURE[t]{
    \epsfig{file=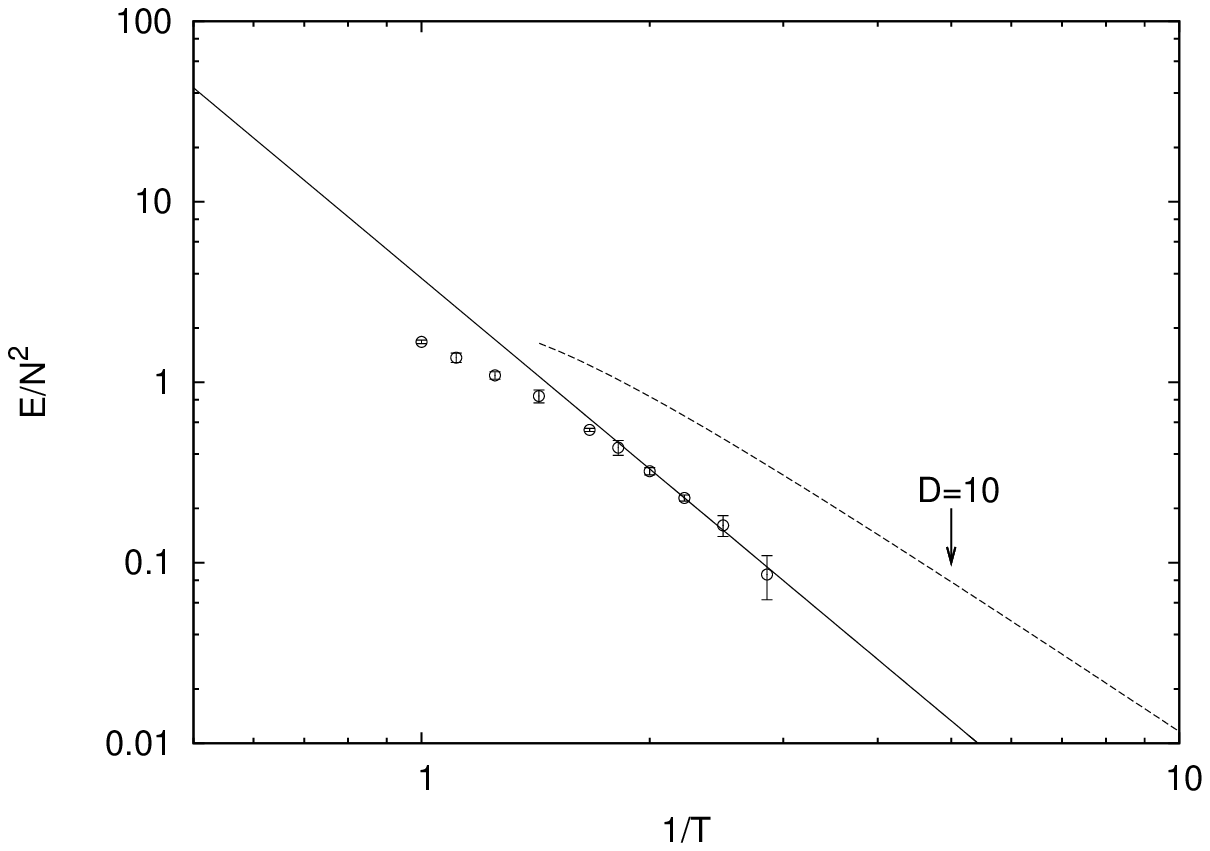,%
width=7.0cm}
    \epsfig{file=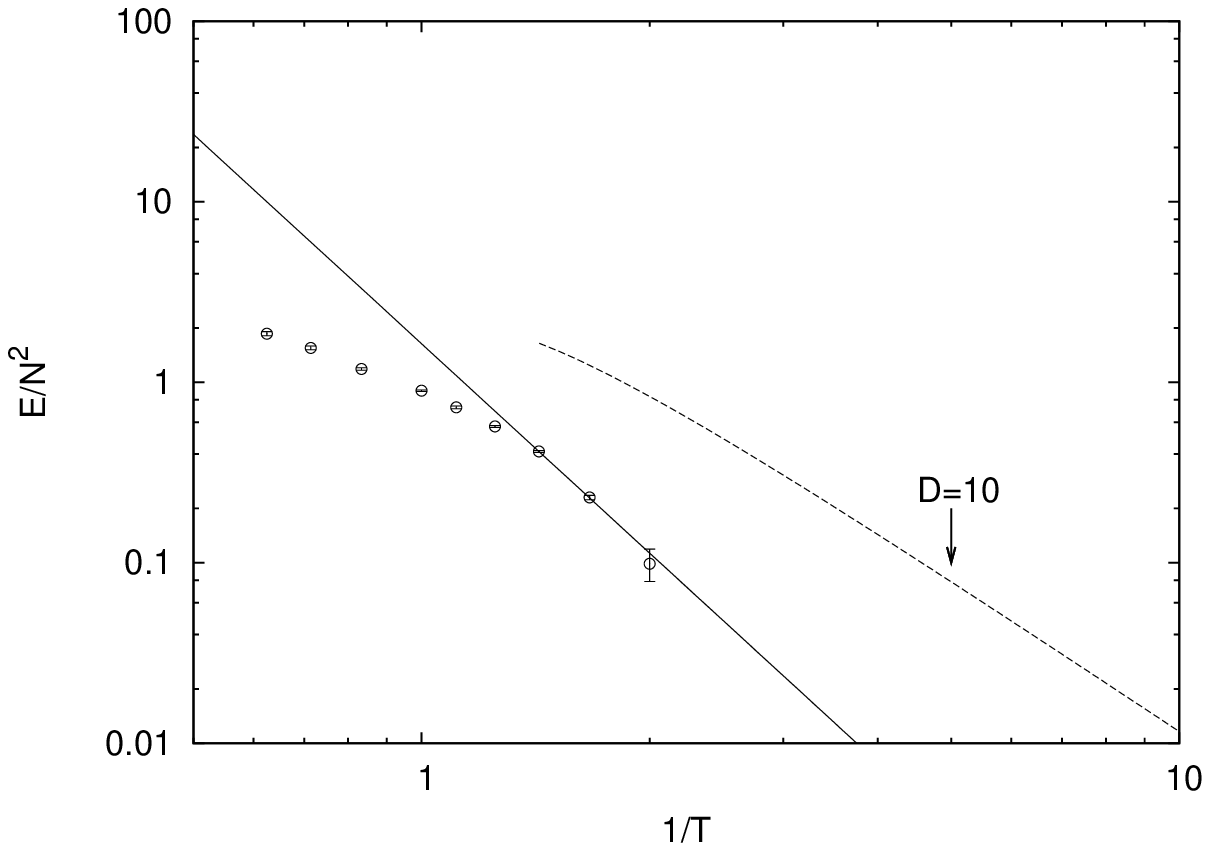,%
width=7.0cm}
\caption{The log-log plot of the
internal energy $\frac{1}{N^2}E$ against the temperature
for the $D=6$ model (left) and for the $D=4$ model (right).
The solid line represents a fit to the power law behavior
(\ref{power-law-behavior})
using the 3 data points with lowest $T$.
The dashed line represents the results for the $D=10$ model \cite{HHNT08}
obtained by fitting the Monte Carlo data to the form
$\frac{1}{N^2}E = 7.41 T^{\frac{14}{5}} 
- c T^{\frac{23}{5}}$, where $c=5.58$.
}
\label{fig:energy_loglog}
}
Fitting the data points with 3 lowest T, we obtain
$a = 5.5 \pm 1.8$, $b = 1.4 \pm 0.1$ for the $D=6$ model,
and $a = 14.1 \pm 0.3$, $b = 2.47 \pm 0.01$
for the $D=4$ model.
The dashed line represents the results for the $D=10$ model,
which cannot be fitted by a straight line at low $T$.
Note also that the internal energy decreases
much more rapidly in the $D=6,4$ models than in the $D=10$ model.

In fig.~\ref{fig:energy_loglog} we show
the log-log plot of the same data. 
A straight line in this figure represents
\beq
\frac{1}{N^2} E = c \, T^q \ ,
\label{power-law-behavior}
\eeq
and the slope is given by $-q$.
Fitting the data points with 3 lowest $T$,
we obtain $c = 3.8 \pm 1.4$, $q = 3.5 \pm 0.5$
for the $D=6$ model 
and $c = 1.63 \pm  0.09$, $q = 3.9 \pm 0.1$
for the $D=4$ model.
The obtained values of $q$ are much larger than 2,
and we do not see any tendency
that the slope becomes closer to $(-2)$ at lower $T$.
Clearly our data are more consistent with (\ref{exp-ansatz})
than with (\ref{power-D46}).

\section{Comparison with the energy spectrum for SU(2)}
\label{sec:su-two}

\FIGURE[H]{
\epsfig{file=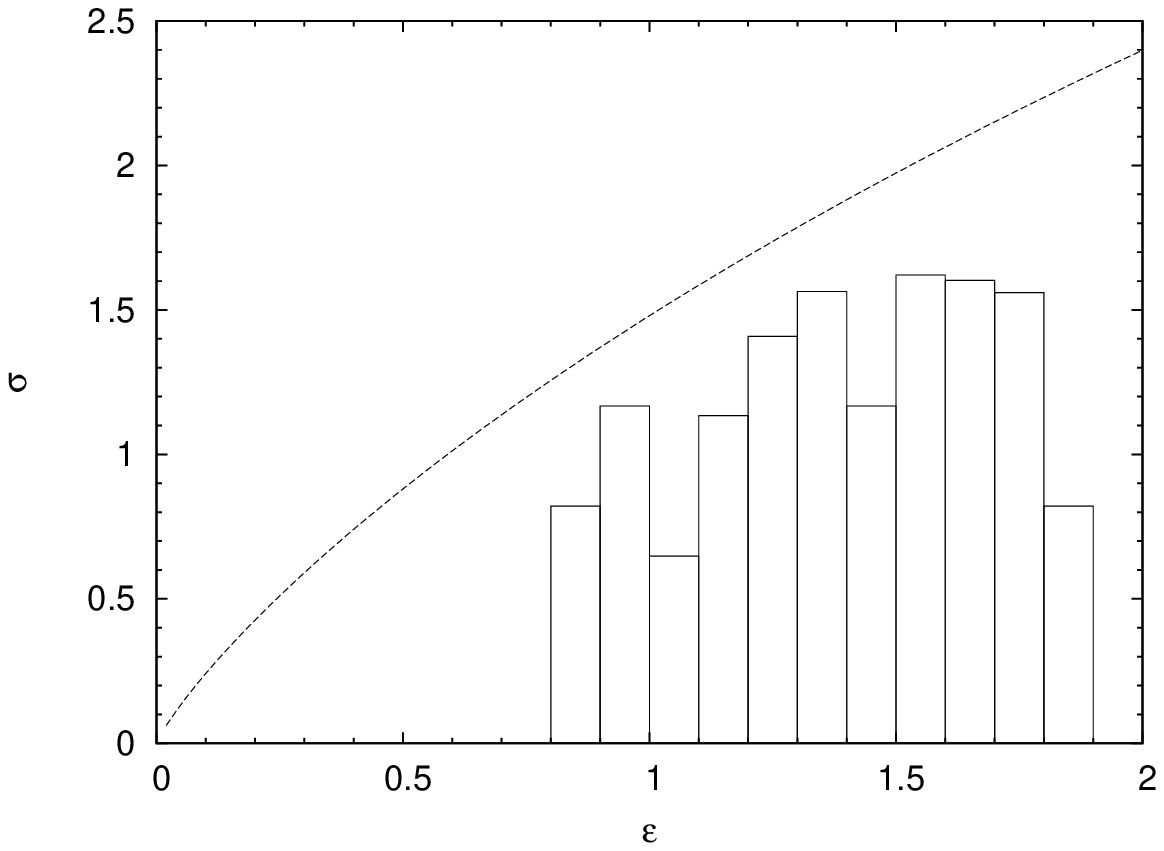,
width=.45\textwidth}
\caption{
The histogram represents
the normalized entropy $\sigma(\varepsilon)$ 
as a function of the normalized energy $\varepsilon$
in the $D=4$ SU(2) supersymmetric model \cite{Campostrini:2004bs}.
%
The dashed line represents the result
(\ref{Ninfty-results})
obtained by fitting the Monte Carlo data for $E(T)$ 
to the behavior (\ref{exp-ansatz-unnorm}).
}
\label{fig:su-two-case}
}

In ref.~\cite{Campostrini:2004bs}
the energy spectrum of 
the $D=4$ model for the gauge group SU(2)
was obtained by diagonalizing the Hamiltonian directly.
The Hilbert space was truncated by the number of bosons.
The discrete states and the continuum states can be clearly 
identified by
the convergence behavior as the cutoff on the number of bosons is increased.
The energy levels for the discrete states
were seen to converge,
whereas those for the continuum states
were seen to decrease slowly without convergence.
The full list of the discrete energy levels 
obtained by this method
is given by tables 9 and 10 of ref.~\cite{Campostrini:2004bs}.
In particular, the lowest level 
in the discrete spectrum is eight-fold 
degenerate.\footnote{Four states form a supermultiplet, 
and there are two of them
due to the symmetry $n_{\rm F} \mapsto 6- n_{\rm F}$, 
where $n_{\rm F}$ represents the number of fermions.
}


On the other hand, from
(\ref{exp-ansatz-unnorm}), 
we obtain 
\beqa
\label{Ninfty-results}
&~& \sigma = A \varepsilon \log \left( \frac{1}{\varepsilon} \right) 
           + B \varepsilon  \nonumber  \ , \\
&~& A=\frac{1}{b} \ , \quad B = \frac{\log a + 1}{b}  \ ,
\label{Ninfty-resultsAB}
\eeqa
where $A=0.405$ and $B=1.48$ plugging in  
the values of $a$ and $b$ obtained
by fitting the Monte Carlo data for $E(T)$ 
to the behavior (\ref{exp-ansatz-unnorm}).

In order to compare the results for $N=2$
with the result (\ref{Ninfty-results}) for $N=\infty$,
we should first note that the coupling constant $g$ is set to 
unity in ref.~\cite{Campostrini:2004bs}.
The internal energy for arbitrary $g$ can be obtained by 
$E(g) = g^{2/3} E(g=1)$.
On the other hand, when we take the large-$N$ limit, we fix $g^2 N=1$.
Therefore, we first rescale the energy spectrum obtained in 
ref.~\cite{Campostrini:2004bs} by a factor of $(\frac{1}{2})^{1/3}$.
Then we count the number of states 
(taking into account the degeneracy) within the interval
$E \sim  E+\delta E$.
Dividing the number by the bin size $\delta E$,
we obtain the density of state $\rho(E)$, and hence
the entropy $S(E)= \log \rho(E)$.
The normalized quantities\footnote{We have also tried
the normalization factor $\frac{1}{N^2-1}$ (instead of $\frac{1}{N^2}$)
motivated by the fact that 
the results in ref.~\cite{Campostrini:2004bs} are obtained
for the gauge group SU(2) instead of U(2). The agreement with the
large-$N$ result does not change drastically, 
however.}
$\sigma \equiv \frac{1}{N^2} S(E)$ and
$\varepsilon \equiv \frac{1}{N^2} E$ are defined
by setting $N=2$.
%
Let us note that $\sigma$ cannot be defined as a continuous function 
of $\varepsilon$ at finite $N$.
In fig.~\ref{fig:su-two-case}
we plot $\sigma$ against $\varepsilon$ as a histogram
with the bin size $\delta \varepsilon = 0.1$.
At larger $N$, it is expected that the histogram becomes smoother
even for smaller bin size, and that one obtains $\sigma$ 
as a continuous function of $\varepsilon$ in the large-$N$ limit.
The dashed line represents 
(\ref{Ninfty-results}) for $N=\infty$ derived from our Monte Carlo results.
It would be interesting to extend the method of 
ref.~\cite{Campostrini:2004bs} to the $N=3$ case,
and to see whether the histogram comes closer
to the $N=\infty$ curve (\ref{Ninfty-results}).


\section{Summary and discussions}
\label{sec:summary}

In this paper we have performed Monte Carlo studies of
supersymmetric matrix quantum mechanics
with 4 and 8 supercharges at finite temperature.
Similarly to the 16 supercharge case studied previously,
the potential for the bosonic matrices 
has a flat direction, which is not lifted quantum
mechanically due to supersymmetry.
This is seen in our Monte Carlo simulation as the run-away behavior.
However, it is possible to suppress this behavior and 
to define a well-defined
ensemble by using sufficiently large $N$.
We consider that
the ensemble obtained by our simulation corresponds to 
restricting 
the Hilbert space 
to normalizable states.

While the phase diagram turns out to be similar to the 16 supercharge
case, we observe a notable difference in the behavior of 
the internal energy. It decreases much faster as the temperature
decreases than in the 16 supercharge case.
In fact, our Monte Carlo data for the internal energy
are consistent 
with exponential decrease at low temperature
in striking contrast to the power-law behavior
in the 16 supercharge case.
%
This behavior was predicted by Smilga assuming
the absence of normalizable states with a new energy scale
unlike in the $D=10$ model.
Thus our results provide independent evidence for
the peculiarity of the $D=10$ model suggested
previously in the literature \cite{Yi:1997eg,Sethi:1997pa,Moore:1998et}.

To our knowledge, there are no concrete proposals for
a gravity dual of the non-maximally supersymmetric models 
studied in the present work.
We hope that our explicit results in the planar large-$N$ limit
would provide a useful guide for constructing such an example.
That would give us new insights into the gauge-gravity duality
in non-maximally supersymmetric cases.

 

\acknowledgments
The authors would like to thank O.~Aharony, B.~Bringoltz, I.~Kanamori,
L.~Mannelli, A.V.~Smilga, M.~\"{U}nsal, J.~Wosiek and L.~Yaffe for useful
discussions and comments.
The work of M.H.\ 
is supported in part 
by JSPS Postdoctoral Fellowship for Research Abroad.
The work of J.N.\ is supported in part by Grant-in-Aid 
for Scientific Research (No.~19340066 and 20540286)
from Japan Society for the Promotion of Science.
S.M.\ would thank the members of Institute of Physics
in Jagiellonian University for their hospitality 
and useful discussion during his stay.
Numerical computations were carried out 
on PC clusters at KEK
and at the Yukawa Institute Computer Facility.

\appendix


\section{Derivation of the formula for the internal energy}
\label{sec:EO}

In this Appendix we explain how we calculate
the internal energy defined by (\ref{defE}).
In ref.~\cite{Kawahara:2007ib}
it is shown that the internal energy $E$
is related to the expectation values
\beq
\label{def E}
\frac{E}{N^2} 
= \langle {\cal E}_{\rm b}  \rangle + 
\langle {\cal E}_{\rm f} \rangle
%
\ ,
\eeq
where the operators 
${\cal E}_{\rm b}$ and ${\cal E}_{\rm f}$
are defined as
\beqa
\label{def-Eb}
{\cal E}_{\rm b} & \equiv &
- \frac{3}{4}\, 
\frac{1}{N\be}
\int_0^\beta  \!\!dt  \, \tr
\Bigl( [X_i,X_j]^2 \Bigr) 
\ ,  \\
{\cal E}_{\rm f} &\equiv& -
\frac{3}{4}\, 
\frac{1}{N\be} 
\int_0^\beta  \!\!dt  \, \tr
\Bigl( \ps_{\alpha} (\gm_i)_{\alpha\beta} [X_i,\ps_{\beta}] \Bigr)
\  .
\label{def-Ef}
\eeqa
However, the calculation of
$\langle {\cal E}_{\rm f} \rangle $ 
is time-consuming since it requires the construction 
of the fermion matrix and inverting it.
In fact we can use an identity to trade it off
with a quantity written solely in terms of bosonic variables.
This trick has been first proposed in ref.~\cite{CW07}
and it was used also in refs.~\cite{AHNT07,HHNT08}.

Let us consider the change of variables
\beq
X_i  \mapsto \ee^{\epsilon} \, X_i (t) 
\eeq
in the path integral representation of the 
partition function (\ref{def-part-fn}).
One finds that the path-integral measure transforms as\footnote{The
``$-1$'' in the exponent is due to the constraint (\ref{czm})}
\beq
{\cal D} X \mapsto \ee^{\epsilon d 
\{ N^2 (2\Lambda+1) - 1 \} } {\cal D} X  \ ,
\eeq
and each term in the action transforms according to 
the order of $X_i$.
Since this is just a change of variables, the value of the partition function
should not change.
This implies that the O($\epsilon$) terms should cancel.
Thus we obtain an identity
\beq
\frac{1}{g^2} \left\langle \intdt {\rm tr}
 \left\{
 2 \oot (D_t X_i)^2 
-  4 \oof [X_i,X_j] ^2
- \oot \ps_\alpha (\gm_i)_{\alpha\beta} [X_i , \ps_\beta ]
\right\} \right\rangle 
= d \{ N^2 (2 \Lambda +1 )-1\} 
\ .
\label{identity-div}
\eeq
Note that the derivative term on the left-hand side
has linear divergence at $\Lambda \rightarrow \infty$,
which is given by the O($\Lambda$) term on the right-hand side.
Using this identity,
we obtain
\beq
\label{final-E}
E
= -\frac{3}{\beta} 
\left[ \langle S_{\rm b}  \rangle 
-\frac{1}{2} d \left\{ (2\Lambda+1) N^2 - 1 \right\} \right]
%
\ .
\eeq
The advantage of using this formula instead of (\ref{def E})
is that the evaluation of $S_{\rm b}$ does not take much time.
The disadvantage is that the fluctuation of $S_{\rm b}$
is O($\sqrt{\Lambda}$), and therefore one needs to increase
the statistics linearly in $\Lambda$ to keep the statistical error
constant. This is related to the fact that the relation
(\ref{final-E}) involves
the cancellation of the O($\Lambda$) terms.
We do not encounter any problem, however,
in the parameter regime investigated in this work.



\begin{thebibliography}{999}

\bibitem{Maldacena97}
J.~M.~Maldacena,
\emph{The large N limit of superconformal field theories 
and supergravity},
\emph{Adv.\ Theor.\ Math.\ Phys.}  {\bf 2} (1998) 231
[\emph{Int.\ J.\ Theor.\ Phys.} {\bf 38} (1999) 1113]
[{\tt hep-th/9711200}].


\bibitem{IMSY98}
  N.~Itzhaki, J.~M.~Maldacena, J.~Sonnenschein and S.~Yankielowicz,
\emph{Supergravity and the large N limit of theories with sixteen
supercharges},
\emph{Phys.\ Rev.\  D} {\bf 58} (1998) 046004 
[{\tt hep-th/9802042}].


\bibitem{HNT07}
  M.~Hanada, J.~Nishimura and S.~Takeuchi,
\emph{Non-lattice simulation for supersymmetric gauge theories in one
dimension},
\emph{Phys.\ Rev.\ Lett.} {\bf 99} (2007) 161602 
[{\tt arXiv:0706.1647}].

\bibitem{AHNT07}
  K.~N.~Anagnostopoulos, M.~Hanada, J.~Nishimura and S.~Takeuchi,
\emph{Monte Carlo studies of supersymmetric matrix quantum mechanics 
with sixteen supercharges at finite temperature},
\emph{Phys.\ Rev.\ Lett.} {\bf 100} (2008) 021601
[{\tt arXv:0707.4454}]. 

\bibitem{HMNT08}
  M.~Hanada, A.~Miwa, J.~Nishimura and S.~Takeuchi,
\emph{Schwarzschild radius from Monte Carlo calculation 
of the Wilson loop in supersymmetric matrix quantum mechanics},
\emph{Phys.\ Rev.\ Lett.} {\bf 102} (2009) 181602
[{\tt arXiv:0811.2081}].

\bibitem{HHNT08}
  M.~Hanada, Y.~Hyakutake, J.~Nishimura and S.~Takeuchi,
\emph{Higher derivative corrections to black hole thermodynamics from
supersymmetric matrix quantum mechanics},
\emph{Phys.\ Rev.\ Lett.} {\bf 102} (2009) 191602
[{\tt arXiv:0811.3102}].

\bibitem{Hanada:2009ne}
  M.~Hanada, J.~Nishimura, Y.~Sekino and T.~Yoneya,
\emph{Monte Carlo studies of Matrix theory correlation functions},
\emph{Phys.\ Rev.\ Lett.} {\bf 104} (2010) 151601
  [arXiv:0911.1623 [hep-th]].


\bibitem{CW07}
  S.~Catterall and T.~Wiseman,
\emph{Towards lattice simulation of the gauge theory 
duals to black holes and hot strings},
\emph{JHEP} {\bf 12} (2007) 104 [{\tt arXiv:0706.3518}].

\bibitem{CW08}
S.~Catterall and T.~Wiseman,
\emph{Black hole thermodynamics from simulations 
of lattice Yang-Mills theory},
\emph{Phys.\ Rev.\ D} {\bf 78} (2008) 041502
[{\tt arXiv:0803.4273}];
%
\emph{Extracting black hole physics from the lattice},
\emph{JHEP} {\bf 1004} (2010) 077
[{\tt arXiv:0909.4947}].

\bibitem{KLL} D.~Kabat, G.~Lifschytz and D.~A.~Lowe, 
\emph{Black hole thermodynamics from calculations 
in strongly coupled gauge theory},
Int.\ J.\ Mod.\ Phys.\ A {\bf 16} (2001) 856 
[\emph{Phys.\ Rev.\ Lett.} {\bf 86} (2001) 1426]
[arXiv:hep-th/0007051]; 
\emph{Black hole entropy from non-perturbative gauge theory},
\emph{Phys.\ Rev.\  D} {\bf 64} (2001) 124015.


\bibitem{BFSS96}
T.~Banks, W.~Fischler, S.~H.~Shenker and L.~Susskind,
\emph{M theory as a matrix model: A conjecture},
\emph{Phys.\ Rev.\  D} {\bf 55} (1997) 5112
[{\tt hep-th/9610043}].

\bibitem{Smilga:1984jg}
  A.~V.~Smilga,
\emph{Witten index calculation in supersymmetric gauge theory},
\emph{Nucl.\ Phys.\  B} {\bf 266} (1986) 45.


\bibitem{de Wit:1988ct}
  B.~de Wit, M.~Luscher and H.~Nicolai,
\emph{The supermembrane is unstable},
\emph{Nucl.\ Phys.\  B} {\bf 320} (1989) 135.

\bibitem{Smilga:1989ew}
  A.~V.~Smilga,
\emph{Super Yang-Mills quantum mechanics and supermembrane spectrum},
\emph{in Proceedings of the Workshop on Supermembranes and $(2+1)$-Dimensional
Physics}, Trieste, 16-23 July 1989, World Scientific, 1990, p.182.



\bibitem{Yi:1997eg}
  P.~Yi,
\emph{Witten index and threshold bound states of D-branes},
\emph{Nucl.\ Phys.\  B} {\bf 505} (1997) 307
[{\tt hep-th/9704098}].

\bibitem{Sethi:1997pa}
  S.~Sethi and M.~Stern,
\emph{D-brane bound states redux},
\emph{Commun.\ Math.\ Phys.} {\bf 194} (1998) 675
[{\tt hep-th/9705046}] . 
%
\bibitem{Moore:1998et}
  G.~W.~Moore, N.~Nekrasov and S.~Shatashvili,
\emph{D-particle bound states and generalized instantons},''
\emph{Commun.\ Math.\ Phys.} {\bf 209} (2000) 77
[{\tt hep-th/9803265}].



\bibitem{Smilga:2008bt}
  A.~V.~Smilga,
\emph{Comments on thermodynamics of supersymmetric matrix models},
\emph{Nucl.\ Phys.\  B} {\bf 818} (2009) 101 
[{\tt arXiv:0812.4753}].






\bibitem{Campostrini:2004bs}
  M.~Campostrini and J.~Wosiek,
\emph{High precision study of the structure 
of D = 4 supersymmetric Yang-Mills quantum mechanics},
\emph{Nucl.\ Phys.\  B} {\bf 703} (2004) 454
[{\tt hep-th/0407021}].

\bibitem{AABHN00}
  J.~Ambjorn, K.~N.~Anagnostopoulos, W.~Bietenholz, T.~Hotta and J.~Nishimura,
  {\it ``Large N dynamics of dimensionally reduced 4D SU(N) super Yang-Mills
   theory,''}
  JHEP {\bf 0007} (2000) 013
  [arXiv:hep-th/0003208].

\bibitem{RHMC}
     M.~ A.~ Clark, and A.~ D.~ Kennedy, and Z.~ Sroczynski,
\emph{Exact 2+1 flavour RHMC simulations},
\emph{Nucl.\ Phys.\ Proc.\ Suppl.} {\bf 140} (2005) 835 
[{\tt hep-lat/0409133}].


\bibitem{Kawahara:2007ib}
  N.~Kawahara, J.~Nishimura and S.~Takeuchi,
\emph{High temperature expansion in supersymmetric matrix quantum mechanics},
\emph{JHEP} {\bf 0712} (2007) 103
[arXiv:0710.2188 [hep-th]].



\bibitem{KS99}
  W.~Krauth and M.~Staudacher,
\emph{Eigenvalue distributions in Yang-Mills integrals},
\emph{Phys.\ Lett.\  B} {\bf 453} (1999) 253 
[{\tt hep-th/9902113}].


\bibitem{latticeBFSS}
R.~A.~Janik and J.~Wosiek,
\emph{Towards the matrix model of M-theory on a lattice},
\emph{Acta Phys.\ Polon.\ B} {\bf 32} (2001) 2143; \\
%
P.~Bialas and J.~Wosiek,
\emph{Towards the lattice study of M-theory (II)},
\emph{Nucl.\ Phys.\ Proc.\ Suppl.} {\bf 106} (2002) 968; \\
%
O.~Aharony, J.~Marsano, S.~Minwalla and T.~Wiseman,
\emph{Black hole - black string phase transitions in thermal 1+1 dimensional
supersymmetric Yang-Mills theory on a circle},
\emph{Class.\ Quant.\ Grav.} {\bf 21} (2004) 5169. \\
%

\bibitem{KNT07}
  N.~Kawahara, J.~Nishimura and S.~Takeuchi,
\emph{Phase structure of matrix quantum mechanics at finite temperature},
\emph{JHEP} {\bf 0710} (2007) 097 
[{\tt arXiv:0706.3517}].


\bibitem{Mandal:2009vz}
  G.~Mandal, M.~Mahato and T.~Morita,
\emph{Phases of one dimensional large N gauge theory in a 1/D expansion},
\emph{JHEP} {\bf 1002} (2010) 034
[{\tt arXiv:0910.4526}].


\bibitem{EK82}
T.~Eguchi and H.~Kawai,
\emph{Reduction of dynamical degrees of freedom in the large N gauge theory},
\emph{Phys.\ Rev.\ Lett.} {\bf 48} (1982) 1063.


\bibitem{NN03}
R.~Narayanan and H.~Neuberger,
\emph{Large N reduction in continuum},
\emph{Phys.\ Rev.\ Lett.} {\bf 91} (2003) 081601
[{\tt hep-lat/0303023}].


\bibitem{Furuuchi:2003sy}
  K.~Furuuchi, E.~Schreiber and G.~W.~Semenoff,
\emph{Five-brane thermodynamics from the matrix model},
{\tt hep-th/0310286}.

\bibitem{Aharony}
O.~Aharony, J.~Marsano, S.~Minwalla, K.~Papadodimas and M.~Van Raamsdonk,
\emph{The Hagedorn / deconfinement phase transition 
in weakly coupled large N gauge theories},
\emph{Adv.\ Theor.\ Math.\ Phys.} {\bf 8} (2004) 603
[\hepth{0310285}].


\bibitem{Becker:1997wh}
  K.~Becker and M.~Becker,
\emph{A two-loop test of M(atrix) theory},
\emph{Nucl.\ Phys.\  B} {\bf 506} (1997) 48
[{\tt hep-th/9705091}].


\bibitem{Okawa:1998pz}
  Y.~Okawa and T.~Yoneya,
\emph{Multi-body interactions of D-particles in supergravity and Matrix
theory},
\emph{Nucl.\ Phys.\  B} {\bf 538} (1999) 67
[{\tt hep-th/9806108}].

\bibitem{Smilga:2002mra}
  A.~V.~Smilga,
\emph{Born-Oppenheimer corrections to the effective 
zero-mode Hamiltonian in SYM theory},
\emph{JHEP} {\bf 0204} (2002) 054
[{\tt hep-th/0201048}].






\end{thebibliography}
\end{document}